\documentclass[epj]{webofc}
\usepackage[utf8]{inputenc}
\usepackage[varg]{txfonts}   
\usepackage{booktabs}
\usepackage{xcolor}
\definecolor{darkred}{rgb}{0.4,0.0,0.0}
\definecolor{darkgreen}{rgb}{0.0,0.4,0.0}
\definecolor{darkblue}{rgb}{0.0,0.0,0.4}
\usepackage[bookmarks,linktocpage,colorlinks,
    linkcolor = darkred,
    urlcolor  = darkblue,
    citecolor = darkgreen]{hyperref}
\wocname{EPJ Web of Conferences}
\woctitle{Lattice2017}
\usepackage{epsfig}
\usepackage{epstopdf}
\newcommand{\bdm}{\begin{dmath}}
\newcommand{\edm}{\end{dmath}}
\newcommand{\bdms}{\begin{dmath*}}
\newcommand{\edms}{\end{dmath*}}
\newcommand{\bdg}{\begin{dgroup*}}
\newcommand{\edg}{\end{dgroup*}}      

\def\lsim{\mathrel{\rlap{\lower4pt\hbox{\hskip1pt$\sim$}}
    \raise1pt\hbox{$<$}}}                
\def\gsim{\mathrel{\rlap{\lower4pt\hbox{\hskip1pt$\sim$}}
    \raise1pt\hbox{$>$}}}                

\def\slashed{{/}\mskip-10.0mu}

\def\circe[#1]{\overset{\circ}{#1}}

\def\openone{\leavevmode\hbox{\small1\kern-3.3pt\normalsize1}}

\def\ZS{Z_{\rm S}}
\def\ZP{Z_{\rm P}}
\def\ZV{Z_{\rm V}}
\def\ZA{Z_{\rm A}}
\def\ZT{Z_{\rm T}}

\newcommand{\be}{\begin{equation}}
\newcommand{\ee}{\end{equation}}
\newcommand{\bea}{\begin{eqnarray}} 
\newcommand{\eea}{\end{eqnarray}}

\newcommand{\Op}{\mathcal{O}} 

\newcommand*\Red[1]{\textcolor{red}{#1}}
\newcommand*\Blue[1]{\textcolor{blue}{#1}}
\newcommand*\Cyan[1]{\textcolor{cyan}{#1}}
\newcommand*\Magenta[1]{\textcolor{magenta}{#1}}

\definecolor{mygreen}{RGB}{30,160,10}

\definecolor{myorange}{RGB}{255,130,0}
\newcommand*\MyOrange[1]{\textcolor{myorange}{#1}}

\begin{document}
%
\selectlanguage{english}
\title{%
Singlet vs Nonsinglet Perturbative Renormalization factors of Staggered Fermion Bilinears
}
\author{%
\firstname{Haralambos} \lastname{Panagopoulos}\inst{1} \and
\firstname{Gregoris}  \lastname{Spanoudes}\inst{1}\fnsep\thanks{Speaker, \email{spanoudes.gregoris@ucy.ac.cy}}
}
\institute{%
Department of Physics, University of Cyprus, POB 20537, 1678, Nicosia, Cyprus
}
\abstract{%
  In this paper we present the perturbative computation of the difference between the renormalization factors of flavor singlet ($\sum_f\bar\psi_f\Gamma\psi_f$, $f$: flavor index) and nonsinglet ($\bar\psi_{f_1} \Gamma \psi_{f_2}, f_1 \neq f_2$) bilinear quark operators (where $\Gamma = \openone,\,\gamma_5,\,\gamma_{\mu},\,\gamma_5\,\gamma_{\mu},\, \gamma_5\,\sigma_{\mu\,\nu}$) on the lattice. The computation is performed to two loops and to lowest order in the lattice spacing, using Symanzik improved gluons and staggered fermions with twice stout-smeared links. The stout smearing procedure is also applied to the definition of bilinear operators. A significant part of this work is the development of a method for treating some new peculiar divergent integrals stemming from the staggered formalism. Our results can be combined with precise simulation results for the renormalization factors of the nonsinglet operators, in order to obtain an estimate of the renormalization factors for the singlet operators. The results have been published in Physical Review D \cite{Constantinou:2016ieh}.
}
\maketitle
\section{Introduction}\label{intro}

Renormalization of flavor singlet operators is essential for the study of a number of hadronic
properties, including topological features and the spin structure
of hadrons; for example, the knowledge of the axial singlet renormalization 
factor is required to compute the light quarks' contribution to the 
spin of the nucleon \cite{QCDSF:2011aa}.
Matrix elements of such operators are notoriously
difficult to study via numerical simulations, due to the presence of
fermion-line-disconnected diagrams, which in principle require
evaluation of the full fermion propagator. Then it is quite a challenge 
to obtain accurate results for the renormalization of the singlet operators 
directly. In recent years there has been some progress in the numerical study 
of flavor singlet operators; for some of them, a nonperturbative 
estimate of their renormalization has been obtained using the Feynman-Hellmann
relation, for both improved Wilson and staggered fermion actions~\cite{Chambers:2014pea,Bouchard:2016heu, Bali:2017jyw}. Perturbation theory can give an important cross
check for these estimates, and provide a prototype for other
operators, such as: $\bar{\psi}\Gamma\,D^\mu\psi$ (appearing in hadron structure
functions) and $(\bar{s}\,\Gamma_1\,d)\,(\bar{s}\,\Gamma_2\,d)$ (appearing in $\Delta
S = 2$ transitions, etc.),which are more difficult to renormalize nonperturbatively. 

Given that the renormalization factors of the nonsinglet 
operators can be calculated nonperturbatively with quite good precision, we can give 
an estimate of the renormalization factors for the singlet operators through the 
perturbative evaluation of the difference between singlet and nonsinglet cases; this 
difference first shows up at two loops. The computation of the two-loop difference between 
the singlet and nonsinglet perturbative renormalization factors of all quark bilinears is the
main goal of this work. Our results are presented in $RI'$ and $\overline{MS}$ renormalization schemes, as well as in an alternative $RI'$ scheme, more appropriate for nonperturbative calculations. Furthermore, we perform the computation, using a class of improved lattice actions: Symanzik improved gluons and staggered fermions with twice stout-smeared links. The corresponding calculation with SLiNC fermions had been previously performed by our group \cite{Constantinou:2014rka}. 

Despite their relatively low computational cost and the absence of additive mass renormalization (due to chiral invariance), staggered fermions entail additional complications in their perturbative study as compared to Wilson fermions. 
In particular, the fact that fermion degrees of freedom are distributed over
neighbouring lattice points requires the introduction of link variables in the
definition of gauge invariant fermion bilinears, with a corresponding increase in the
number of Feynman diagrams. In addition, the appearance of 16 (rather than 1) poles
in the fermion propagator leads to a rather intricate structure of divergent contributions 
in two-loop diagrams. 

A novel aspect of the calculation is that the gluon links, which appear both in the staggered fermion action and in the definition of the staggered bilinear operators, are improved by applying a stout smearing procedure up to two times, iteratively. Compared to most other improved formulations of staggered fermions, the stout smearing action leads to smaller taste violating effects \cite{Aoki:2005vt,Borsanyi:2011bm,Bazavov:2012zad}. However, double stout smearing produces an enormous number of terms to evaluate. Application of stout improvement on staggered fermions thus far has been explored, by our group, only to one-loop computations \cite{Constantinou:2013pba}; a two-loop computation had never been investigated before. 

\section{Formulation and Calculational Setup}\label{sec-2}

\subsection{Lattice actions}\label{subsec-2.1}

In our calculation we made use of the staggered formulation 
of the fermion action on the lattice, applying a twice stout smearing procedure on the gluon links. 
In standard notation, it reads:  
\be
S_{\rm SF} = a^4 \sum_{x,\mu} \frac{1}{2a} \, \overline{\chi} (x) 
\ \eta_\mu (x) \ \Big[ \widetilde{\widetilde{U}}_\mu (x) \ \chi (x + a \hat{\mu})
- \widetilde{\widetilde{U}}_\mu^\dagger (x - a \hat{\mu}) \ \chi (x - a \hat{\mu}) \Big] + a^4 \sum_{x} m \ \overline{\chi} (x) \ \chi (x)\,,
\label{SFactionimpr1}
\ee
where $\chi (x)$ is a one-component fermion field, and $\eta_\mu (x) = (-1)^{\sum_{\nu < \mu} n_\nu}$ 
[$ x =(a\,n_1,a\,n_2,a\,n_3,a\,n_4),\quad n_i\,\, \epsilon\,\, {\mathbb Z}\,$]. 
The relation between the staggered field $\chi(x)$ 
and the standard fermion field $\psi (x)$, is given by: 
$\psi(x) = \gamma_x\,\chi(x),\quad \bar\psi(x) = \bar\chi(x)\,\gamma_x^\dagger\,$,
where $\gamma_x=\gamma_1^{n_1}\,\gamma_2^{n_2}\,\gamma_3^{n_3}\,\gamma_4^{n_4}$.
Since a single fermion field component $\chi (x)$ corresponds to each lattice site, the staggered action contains 4 rather than 16 fermion doublers, which are called ``tastes''. Then, a physical fermion field $\psi (x)$ with taste components (totally 16 components) lives in a 4-dimensional unit hypercube of the lattice. The gluon links 
$\widetilde{\widetilde{U}}_\mu (x)$, appearing above, 
are ``doubly'' stout links, defined as:
$\widetilde{\widetilde{U}}_\mu(x) = e^{i\,\widetilde{Q}_\mu(x)}\,\widetilde{U}_\mu(x)\,$,
where $\widetilde{U}_\mu(x)$ is the ``singly'' stout link \cite{Morningstar:2003gk}:
\begin{gather}
\widetilde{U}_\mu(x) = e^{i\,Q_\mu(x)}\,U_\mu(x)\,, \ Q_\mu(x)= \frac{\omega}{2\,i} \Big[V_\mu(x) U_\mu^\dagger(x) - U_\mu(x) V_\mu^\dagger(x) -\frac{1}{N_c} {\rm Tr} \,\Big(V_\mu(x) U_\mu^\dagger(x) -  U_\mu(x) V_\mu^\dagger(x)\Big)\Big]\,, \nonumber \\
V_{\mu} (x) = \sum_{\rho = \pm 1}^{\pm 4} U_{\rho} (x) U_{\mu} (x + a \hat{\rho}) U_{\rho}^{\dagger} (x + a \hat{\mu})\,.
\label{Ustout_def}
\end{gather}
$V_\mu(x)$ represents the sum over all staples associated with the
link $U_\mu(x)$, $\omega$ is a tunable parameter, called stout smearing parameter and $N_c$ is the number of colors. 
Correspondingly, $\widetilde{Q}_\mu(x)$ is defined as in Eq.\eqref{Ustout_def}, but using 
$\widetilde{U}_\mu$ as links (also in the construction of $V_\mu$). 
To obtain results 
that are as general as possible, we use different stout parameters, $\omega$, in the first 
($\omega_1$) and the second ($\omega_2$) smearing iteration. 

For gluons, we employ a Symanzik improved action, of the form \cite{Horsley:2004mx}:
\be
S_G=\frac{2}{g_0^2} {\rm Re\,Tr\,} \Bigl[ c_0 \sum_{\rm plaq.} (1-U_{\rm plaq.})\,+\, c_1 \sum_{\rm rect.} (1- U_{\rm rect.}) + c_2 \sum_{\rm chair} (1-U_{\rm chair})\,+\, c_3 \sum_{\rm paral.} (1-U_{\rm paral.})\Bigr]\,,
\label{Symanzik}
\ee
where $U_{\rm plaq.}$ is the 4-link Wilson loop (``$1 \times 1$ plaquette'') and $U_{\rm rect.}$, $U_{\rm chair}$, 
$U_{\rm paral.}$ are the three possible independent 6-link Wilson loops (``$2 \times 1$ rectangle'', ``$2 \times 1$ chair'', ``$2 \times 1$ parallelogram'').
The Symanzik coefficients $c_i$ satisfy the normalization condition: $c_0 + 8 c_1 + 16 c_2 + 8 c_3 = 1\,$.
We have selected a number of commonly used sets of values for $c_i$, some of which are shown in Table \ref{tab1}.
\begin{table}[thb]
  \centering
  \begin{tabular}{lllll}\toprule
  Gluon action  & $c_0$ & \quad $c_1$ & \quad $c_2$ & \quad $c_3$  \\\midrule
  Wilson & \ $1$ & \quad $0$ & \quad $0$ & \quad $0$ \\
  Tree-Level Symanzik & $5/3$ & $-1/12$ & \quad $0$ & \quad $0$ \\
  Iwasaki & $3.648$ & $-0.331$ & \quad $0$ & \quad $0$ \\\bottomrule
  \end{tabular}
  \caption{Selected sets of values for Symanzik coefficients.}
\label{tab1}
\end{table}

\subsection{Definition of Staggered fermion bilinear operators}\label{subsec-2.2}

In the staggered formalism, each physical fermion field component $\psi_{\alpha, t}$ (where $\ \alpha$ is a Dirac index and $t$ is a taste index) is defined as a linear combination of the single-component fermion fields $\chi$ that live on the corners of 4-dimensional elementary hypercubes of the lattice. In standard notation:
\be
\psi_{\alpha,t}(y) = \frac{1}{2}\,\sum_C\left(\gamma_C\right)_{\alpha,t}\,\chi_C (y)\,,\quad
\chi_C (y) = \frac{1}{2} \sum_{\alpha,t}\left(\xi_C\right)_{\alpha,t}\,\psi_{\alpha,t}(y)
\label{Stagg_rot}
\ee
where $\chi_C (y)\equiv\chi(y+aC)/4$, $y$ denotes the position of a hypercube inside the lattice ($y_\mu\,\in\,\,2{\mathbb Z}$), C denotes the position of a fermion field component within a specific hypercube ($C_\mu \in \{0,1\}$), $\gamma_C = \gamma_1^{C_1}\,\gamma_2^{C_2}\,\gamma_3^{C_3}\,\gamma_4^{C_4}$, $\ \xi_C = \xi_1^{C_1}\,\xi_2^{C_2}\,\xi_3^{C_3}\,\xi_4^{C_4}$ and $\ \xi_\mu = \gamma_\mu^\ast$.

Using Eq.\eqref{Stagg_rot}, one can define the fermion bilinear operators ${\cal O}_{\Gamma,\Xi}=\bar \psi (y)\,\left(\Gamma\otimes\Xi\right)\,\psi (y)\,$ (where $\Gamma$ and $\Xi$ are arbitrary $4\times4$ matrices acting on the Dirac and taste indices of $\psi_{\alpha,t} (y)$, respectively) in terms of the staggered fermion fields $\chi_C (y)$ \cite{Patel:1992vu}:
\be
{\cal O}_{\Gamma,\Xi} = \sum_{C,D}
\bar\chi_C (y)\,\frac{1}{4}\,{\rm
  Tr}\left[\gamma^\dagger_C\,\Gamma\,\gamma_D\,\Xi\right]\,U_{C,D}\,\chi_D (y)\,,
\label{O_general}
\ee
where one inserts the quantity $U_{C,D}$, which is the average of products of gauge link variables along all possible shortest paths connecting the sites $y+C$ and $y+D$, in order to restore gauge invariance. Using the relations $\gamma_\mu \gamma_C = \eta_\mu (C) \gamma_{C + \hat{\mu}}$ and $\rm{Tr}$ $( \gamma_C^\dagger \gamma_D ) = 4 \delta_{C,D}$, the taste-singlet staggered fermion bilinear operators take the following form:
\begin{gather}
{\cal O}_S(y) = \sum_D \bar\chi_D (y) \, \chi_D (y)\,, \quad {\cal O}_V(y) = \sum_D \bar\chi_{D+_{_2}\hat{\mu}} (y)\,U_{D+_{_2}\hat{\mu},D}\,\chi_D (y)\,\eta_\mu(D)\,, 
\label{O_V}\\
{\cal O}_T(y) = \frac{1}{i} \sum_D\,\bar\chi_{D+_{_2}\hat{\mu}+_{_2}\hat{\nu}} (y)\,U_{D+_{_2}\hat{\mu}+_{_2}\hat{\nu},D} \,\chi_D (y) \,\eta_\nu(D) \,\eta_{\mu}(D+_{_2}\hat{\nu})\,,\ \mu \neq \nu\,, 
\label{O_T}\\
{\cal O}_P(y) = \sum_D \bar\chi_{D+_{_2}(1,1,1,1)} (y)\,U_{D+_{_2}(1,1,1,1),D}\,\chi_D (y)\,\eta_1(D)\,\eta_2(D)\,\eta_3(D)\,\eta_4(D)\,,
\label{O_P}
\end{gather}
\begin{gather}
{\cal O}_A(y) = \sum_D\bar\chi_{D+_{_2}\hat{\mu}+_{_2}(1,1,1,1)} (y)\,U_{D+_{_2}\hat{\mu}+_{_2}(1,1,1,1),D}\,\chi_D (y)\,\eta_\mu(D)\, \eta_1(D+_{_2}\hat{\mu})\,\eta_2(D+_{_2}\hat{\mu})\,\eta_3(D+_{_2}\hat{\mu})\,\eta_4(D+_{_2}\hat{\mu})\,,
\label{O_A}
\end{gather}
where $a +_{_2} b \equiv (a+b)$ mod $2$ and $S (\rm Scalar)$, $P (\rm Pseudoscalar)$, $V (\rm Vector)$, $A (\rm Axial \ Vector)$, $T (\rm Tensor)$ correspond to: $\Gamma = \openone, \gamma_5, \gamma_\mu, \gamma_5 \gamma_\mu, \gamma_5 \ \sigma_{\mu \nu}$ respectively and $\Xi = \openone$.

Just as in the staggered fermion action, the gluon links used in the operators, are doubly stout links. We have kept the stout parameters of the action (${\omega}_{A_1},{\omega}_{A_2}$) distinct from the stout parameters of the operators (${\omega}_{O_1},{\omega}_{O_2}$), for wider applicability of the results.

\subsection{Renormalization of fermion bilinear operators}\label{subsec-2.3}

The renormalization factors $Z_\Gamma$ for lattice fermion bilinear 
operators relate the bare operators 
$\mathcal{O}_{\Gamma_\circ} = \bar{\psi} \Gamma \psi$ to their 
corresponding renormalized continuum operators 
$\mathcal{O}_\Gamma$ via: $\mathcal{O}_\Gamma = Z_\Gamma \,\mathcal{O}_{\Gamma_\circ}$.
In order to calculate the renormalization 
factors $Z_\Gamma$, it is essential to compute the 2-point amputated Green's functions of the operators $\mathcal{O}_{\Gamma_\circ}$; they can be written in the following form:
\begin{gather}
\Sigma_{\rm S} (a q) = \openone \ \Sigma^{(1)}_{\rm S} (a q), \quad \Sigma_{\rm P} (a q) = \gamma_5 \ \Sigma^{(1)}_{\rm P} (a q) \label{GreensFunctionSP}\\
\Sigma_{\rm V} (a q) = \gamma_{\mu} \ \Sigma^{(1)}_{\rm V} (a q) +  \frac{q^{\mu} \slashed{q}}{q^2} \ \Sigma^{(2)}_{\rm V} (a q), \quad \Sigma_{\rm A} (a q) = \gamma_5 \ \gamma_{\mu} \ \Sigma^{(1)}_{\rm A} (a q) + \gamma_5 \ \frac{q^{\mu} \slashed{q}}{q^2} \ \Sigma^{(2)}_{\textrm A } (a q) \label{GreensFunctionVA}\\
\Sigma_{\textrm T} (a q) = \gamma_5 \ \sigma_{\mu \nu} \ \Sigma^{(1)}_{\rm T} (a q) + \gamma_5 \ \frac{\slashed{q}}{q^2} (\gamma_{\mu} q_{\nu} - \gamma_{\nu} q_{\mu}) \ \Sigma^{(2)}_{\rm T} (a  q) \label{GreensFunctionT}
\end{gather}
where $\Sigma^{(1)}_{\Gamma} = 1 + \mathcal{O}(g_{\circ}^2), \ \Sigma^{(2)}_\Gamma= {\cal O}(g_\circ^2)$, $g_\circ$: bare coupling constant. 

The renormalization condition in $RI'$ scheme giving $Z_\Gamma^{L,RI'}$ (L: Lattice regularization) is:
\be
\lim_{a \rightarrow 0} \bigg[ Z_{\psi}^{L,RI'} \ Z_{\Gamma}^{L,RI'} \ \Sigma_{\Gamma}^{(1)} (a q) \bigg]_{\begin {smallmatrix}
\ q^2 = {\bar{\mu}}^2, \\
m = 0
\end{smallmatrix}}= 1
\label{ZGamma}
\ee
where $\bar{\mu}$ is the renormalization scale (normally chosen equal to $\overline{MS}$ scale, i.e. $\bar{\mu} = \mu \ (4 \pi / e^{\gamma_E})^{1/2}$) and $Z_\psi$ is the renormalization factor for the fermion field ($\psi = Z_\psi^{-1 / 2} \ \psi_\circ$, $\psi (\psi_\circ)$: renormalized (bare) fermion field). 
This scheme does not involve 
$\Sigma_\Gamma^{(2)}$; nevertheless, renormalizability of the theory implies that $Z_\Gamma^{L,RI'}$ will render the entire Green's function finite.

An alternative prescription of $RI'$ scheme has the following renormalization condition for $Z_\Gamma$:
\be
\lim_{a \rightarrow 0} \bigg[ Z_{\psi}^{L,RI'} Z_{\Gamma}^{L,RI'(\text{alter})} \ \frac{\text{tr} \big( \Gamma \Sigma_{\Gamma} (a q) \big)}{\text{tr} \big( \Gamma \Gamma \big)} \bigg]_{\begin {smallmatrix}
\ q^2 = {\bar{\mu}}^2, \\
m = 0
\end{smallmatrix}}= 1
\ee
where a summation over repeated indices $\mu$ and $\nu$ is understood. This scheme has the advantage of taking into account the whole bare Green's function and therefore is more appropriate for nonperturbative renormalization via numerical simulations where the arithmetic data for $\Sigma_\Gamma$ cannot be separated into two different structures. $RI'$ and $RI'$-alternative prescriptions differ between themselves (for V, A, T) by a finite amount.

The renormalization factors $Z_\Gamma^{L,\overline{MS}}$ for the operators $\mathcal{O}_{\Gamma_\circ}$ in $\overline{MS}$ scheme can be evaluated using the regularization independent conversion factors $C_\Gamma (g, \alpha)$ between $RI'$ and $\overline{MS}$ schemes, as below:
\be
Z_\Gamma^{L,\overline{MS}} = Z_\Gamma^{L,RI'} / C_\Gamma \quad (\textrm{for} \quad \Gamma= S, V, T) , \quad  Z_{\rm P}^{L,\overline{MS}} = Z_{\rm P}^{L,RI'} / C_{\rm S} Z_5^{\rm P}, \quad Z_{\rm A}^{L,\overline{MS}} = Z_{\rm A}^{L,RI'} / C_{\rm V} Z_5^{\rm A} \label{ZPA}
\ee
where $Z_5^{\rm P} (g)$ and $Z_5^{\rm A} (g)$ are additional finite factors, so that Pseudoscalar and Axial Vector operators satisfy the Ward identities; we also note that the value of $Z_5^A$ for the flavor singlet operator differs from that of the nonsinglet one. 
The values of the conversion and ``$Z_5$'' factors are calculated in Refs. \cite{Gracey:2003yr, Larin:1993tq}.

\section{Computation and Results}\label{sec-3}

\subsection{Feynman Diagrams}\label{subsec-3.1}

\begin{figure}[htbp]
\centering
\includegraphics[width=8.5cm]{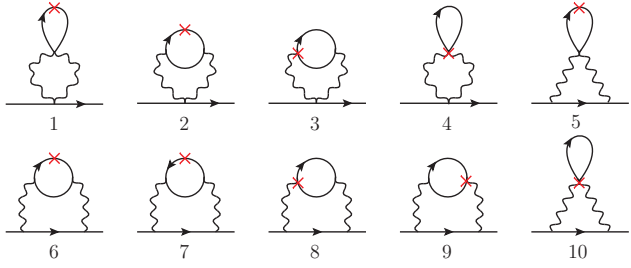}
\caption{Diagrams contributing to the difference between flavor singlet and nonsinglet values of $Z_{\Gamma}$. Solid (wavy) lines represent fermions (gluons). A cross denotes insertion of the operator $\mathcal{O}_{\Gamma}$.}
\label{figDiagS}
\end{figure}

There are 10 two-loop Feynman diagrams that enter in the computation of the 2-point amputated Green's functions of the operators, contributing to the difference between flavor singlet and nonsinglet operator renormalization; they are shown in Fig.  ~\ref{figDiagS}. They all contain an operator insertion inside a closed fermion loop, and therefore vanish in the flavor nonsinglet case. Given that this difference first arises at two loops, we only need the tree-level values of $Z_\psi$ and of the conversion factors $C_\Gamma$, $Z_5^P$. 

The contribution of the diagrams in Fig. \ref{figDiagS} to $\ZP$, $\ZV$, $\ZT$ vanishes identically just as in continuum regularizations. Unlike the case of Wilson fermions \cite{Constantinou:2014rka}, $Z_S$ also vanishes for staggered fermions. The closed fermion loop of the diagrams which contribute to $\ZS$, $\ZP$, $\ZT$, gives an odd number of exponentials of the inner momentum; this leads to odd integrands, which equal zero, due to the symmetry $p_\mu \rightarrow p_\mu + \pi \hat{\nu}$ (where $\mu,\nu$ can be in the same or in different directions) of the staggered propagator. So, for the cases of $\ZS$, $\ZP$, $\ZT$, the contribution vanishes diagram by diagram. Conversely, for the case of $\ZV$, each diagram vanishes when we add its symmetric diagram (diagrams 6+7, 8+9). Therefore, only $\ZA$ is affected; in particular, only diagrams 6 - 9 contribute to $\ZA$.

\subsection{Technical aspects: Treatment of nontrivial divergent integrals}\label{subsec-3.2}

The different pole structure of the staggered fermion propagator gives rise to some nontrivial divergent integrals in the computation of the above two-loop diagrams. In particular, there appeared 4 types of nontrivial divergent 2-loop integrals: 
\begin{small}
\begin{gather}
{I_1}_{\mu \nu} = \int_{- \pi}^{\pi} \frac{d^4 k}{{(2 \pi)}^4} \frac{\circe[k]_{\mu} \ \circe[k]_{\nu}}{(\widehat{k}^2)^2 \  (\widehat{k + a q})^2} \int_{- \pi}^{\pi} \frac{d^4 p}{{(2 \pi)}^4} \frac{1}{\circe[p^2] \ (\circe[p + k])^2}, \ {I_2}_{\mu \nu} = \int_{- \pi}^{\pi} \frac{d^4 k}{{(2 \pi)}^4} \frac{\circe[k]_{\mu} \ \circe[(a q)]_{\nu}}{(\widehat{k}^2)^2 \ (\widehat{k + a q})^2} \int_{- \pi}^{\pi} \frac{d^4 p}{{(2 \pi)}^4} \frac{1}{\circe[p^2] \ (\circe[p + k])^2}, \label{Int12} \\
{I_3}_{\mu \nu \rho \sigma} = \int_{- \pi}^{\pi} \frac{d^4 k}{{(2 \pi)}^4} \frac{\circe[k]_{\mu} \ \circe[k]_{\nu}}{(\widehat{k}^2)^2 \  (\widehat{k + a q})^2} \int_{- \pi}^{\pi} \frac{d^4 p}{{(2 \pi)}^4} \frac{\circe[(2 p)]_{\rho} \ \circe[(2 p)]_{\sigma}}{(\circe[p^2])^2 \ (\circe[p + k])^2}, \ {I_4}_{\mu \nu \rho \sigma} = \int_{- \pi}^{\pi} \frac{d^4 k}{{(2 \pi)}^4} \frac{\circe[k]_{\mu} \ \circe[(a q)]_{\nu}}{(\widehat{k}^2)^2 \ (\widehat{k + a q})^2} \int_{- \pi}^{\pi} \frac{d^4 p}{{(2 \pi)}^4} \frac{\circe[(2 p)]_{\rho} \ \circe[(2 p)]_{\sigma}}{(\circe[p^2])^2 \ (\circe[p + k])^2},\label{Int34}
\end{gather}
\end{small} 
where $\widehat{r}^2 = \sum_{\mu} \widehat{r}^2_\mu$, \ $\widehat{r}_\mu = 2 \sin(r_\mu / 2)$, \ $\circe[r]^2 = \sum_\mu \circe[r]^2_\mu$, \ $\circe[r]_\mu = \sin(r_\mu)$, (in this case $r = p$ or $k$ or $aq$ or $2p$ or $(p+k)$ or $(k+aq)$) and q is an external momentum. The crucial point is the fact that we cannot apply standard subtractions of the form $1 / \circe[p]^2 = 1 / \overset{}{\widehat{p}}^2 + \Big( 1 / \circe[p]^2 - 1 / \overset{}{\widehat{p}}^2 \Big)$, as in Wilson fermions, because of the existence of potential IR singularities at all corners of the Brillouin zone (not only at zero momentum), in the staggered fermion propagator. Therefore, such a subtraction will not alleviate the divergent behaviour at the remaining corners of the Brillouin zone. 

Below, we describe a proposed method for treating such nontrivial integrals. At first, we perform the substitution $p_{\mu} \rightarrow p_{\mu}' + \pi \ C_\mu$, where $- \pi/2 < p_\mu' < \pi / 2$ and $C_\mu \in \lbrace 0,1 \rbrace$. Now the integration region for the innermost integral breaks up into 16 regions with range $[- \pi/2, \pi / 2]$; the contributions from these regions are identical. To restore the initial range $[- \pi, \pi]$, we apply the following change of variables: $p_\mu' \rightarrow p_\mu'' = 2 p_\mu'$. Next, we apply subtractions of the form: $A(2k) = A(k) + [A_{\rm as} (2k) - A_{\rm as} (k)] + [A(2k) - A(k) - A_{\rm as} (2k) + A_{\rm as} (k)]$ and $B_{\rho \sigma} (2k) = \widetilde{B}_{\rho \sigma} (2k) + [B_{\rho \sigma} (2k) - \widetilde{B}_{\rho \sigma} (2k)]$, where
\begin{small}
\bea
A(k) &=& \int_{- \pi}^{\pi} \frac{d^4 p}{{(2 \pi)}^4} \frac{1}{\widehat{p}^2 \ (\widehat{p + k})^2}, \quad A_{\rm as} (k) \equiv \frac{1}{(4 \pi)^2}[- \ln(k^2) + 2] + P_2 \label{A_k}\\
B_{\rho \sigma} (k) &=& \int_{- \pi}^{\pi} \frac{d^4 p}{{(2 \pi)}^4} \frac{\circe[p]_{\rho} \ \circe[p]_{\sigma}}{(\widehat{p}^2)^2 \  (\widehat{p + k})^2}, \quad \widetilde{B}_{\rho \sigma} (2k) \equiv \frac{1}{2 (4 \pi)^2} \frac{\circe[k]_{\rho} \ \circe[k]_{\sigma}}{\widehat{k}^2} + \delta_{\rho \sigma} [\frac{1}{4} A(2k) - \frac{1}{32} P_1] \label{B_k}
\eea
\end{small}
Then, we end up with standard (in the literature) divergent integrals \cite{Luscher:1995np, Chetyrkin:1981, Panagopoulos:1990} and convergent terms that we can integrate numerically for $a \rightarrow 0$. The final expressions for the four integrals are given by:
\begin{small}
\bea
{I_1}_{\mu \nu} &=& \Big\lbrace \frac{2}{(2 \pi)^4} \Big[ - \ln (a^2 q^2) + \frac{3}{2} - \ln 4 \Big] + \frac{1}{2 \pi^2} P_2 \Big\rbrace \frac{q_\mu q_\nu}{q^2} + \delta_{\mu \nu} \Big\lbrace \frac{2}{(4 \pi)^4} \Big[ \ln(a^2 q^2) \Big]^2 \nonumber \\
&-& \frac{1}{4 \pi^2} \Big[P_2 + \frac{1}{(4 \pi)^2} \Big( \frac{5}{2} - \ln 4 \Big) \Big] \ln (a^2 q^2) - \frac{1}{4 \pi^2} \Big[ P_2 + \frac{3}{2 (4 \pi)^2} \ln 4 \Big] + 4 X_2 + \color{blue}G_1 \color{black}\Big\rbrace + \mathcal{O} ( a^2 q^2) \label{I1} \\
{I_2}_{\mu \nu} &=& \Big\lbrace \frac{1}{(2 \pi)^4} \Big[ \ln (a^2 q^2) - 2 + \ln 4 \Big] - \frac{1}{\pi^2} P_2 \Big\rbrace \frac{q_\mu q_\nu}{q^2} + \mathcal{O} ( a^2 q^2) \label{I2}\\
{I_3}_{\mu \nu \rho \sigma} &=& \frac{1}{3 (2 \pi)^4} \frac{q_\mu q_\nu q_\rho q_\sigma}{q^4} + \delta_{\rho \sigma} \Big\lbrace \frac{2}{(2 \pi)^4} \Big[ - \ln (a^2 q^2) + \frac{5}{3} - \ln 4 \Big] - \frac{1}{(4 \pi)^2} (P_1 - 8 P_2) \Big\rbrace \frac{q_\mu q_\nu}{q^2} \nonumber \\
&+& \frac{1}{12 (2 \pi)^4} \Big\lbrace \delta_{\mu \nu} \frac{q_\rho q_\sigma}{q^2} + \delta_{\mu \rho} \frac{q_\nu q_\sigma}{q^2} + \delta_{\mu \sigma} \frac{q_\nu q_\rho}{q^2} + \delta_{\nu \rho} \frac{q_\mu q_\sigma}{q^2} + \delta_{\nu \sigma} \frac{q_\mu q_\rho}{q^2} \Big\rbrace + \delta_{\mu \nu} \delta_{\rho \sigma} \Big\lbrace \frac{2}{(4 \pi)^4} \Big[ \ln(a^2 q^2) \Big]^2 \nonumber \\
&-& \frac{1}{4 \pi^2} \Big[P_2 - \frac{1}{8} P_1 + \frac{1}{(4 \pi)^2} \Big( \frac{51}{2} - \ln 4 \Big) \Big] \ln (a^2 q^2) - \frac{1}{4 \pi^2} \Big[ \Big( \frac{1}{3} - \ln 4 \Big) P_2 - \frac{11}{144} P_1 \nonumber \\
&+& \frac{3}{2 (4 \pi)^2} \Big( \frac{1}{27} - \ln 4 \Big) \Big] - \frac{1}{2} P_1 \ P_2 + 4 X_2 + \color{blue} G_1 \color{black} + \color{blue} G_3 \color{black}\Big\rbrace + (\delta_{\mu \rho} \delta_{\nu \sigma} + \delta_{\mu \sigma} \delta_{\nu \rho}) \Big\lbrace \frac{1}{(12 \pi)^4} \Big[ - \ln (a^2 q^2) \nonumber \\
&+& \frac{1}{6}  \Big] + \frac{1}{6 \pi^2} (P_1 + 3 P_2) + \color{blue} G_2 \color{black}\Big\rbrace + \delta_{\mu \nu \rho \sigma} \Big\lbrace \frac{1}{(2 \pi)^4} + \frac{1}{2 (4 \pi)^2} - \frac{1}{3 \pi^2} P_1 + \color{blue} G_4 \color{black} \Big\rbrace + \mathcal{O} ( a^2 q^2) \label{I3} \\
{I_4}_{\mu \nu \rho \sigma} &=& - \frac{1}{2 (2 \pi)^4} \frac{q_\mu q_\nu q_\rho q_\sigma}{q^4} - \frac{4}{(4 \pi)^4} \Big\lbrace \delta_{\mu \rho} \frac{q_\nu q_\sigma}{q^2} + \delta_{\mu \sigma} \frac{q_\nu q_\rho}{q^2} \Big\rbrace + \delta_{\rho \sigma} \Big\lbrace \frac{1}{(2 \pi)^4} \Big[ \ln (a^2 q^2) - \frac{9}{4} \Big] \nonumber \\
&-& \frac{1}{2 (2 \pi)^2} (P_1 - 8 P_2) \Big\rbrace \frac{q_\mu q_\nu}{q^2} + \mathcal{O} ( a^2 q^2) \label{I4}
\eea
\end{small}
where $P_1, P_2, X_2$ are given in Ref. \cite{Luscher:1995np} and \color{blue} $G_1$ \color{black} $= 0.000803016(6)$, \color{blue} $G_2$ \color{black} $= -0.0006855532(7)$, \color{blue} $G_3$ \color{black} $= 0.00098640(7)$ and \color{blue} $G_4$ \color{black} $= 0.00150252(2)$.

\subsection{Final results}\label{subsec-3.3}

As we have noted in section 3.1, the two-loop results for Scalar, Pseudoscalar, Vector and Tensor operators are zero; for the Axial Vector operator, our result can be written in the following form: 
\vspace{-0.25cm}
\begin{small}
\bea 
&& \hspace{-1.5cm}\ZA^{\rm RI' (singlet)} (a \bar{\mu}) - \ZA^{\rm RI' (nonsinglet)} (a \bar{\mu}) = \nonumber \\
&& \hspace{-1.5cm}- \ \frac{g_\circ^4}{(4 \pi)^4} \ c_F \ N_f \ \Bigg\lbrace \ \MyOrange{\mathbf{6} \ \boldsymbol{\ln}(\boldsymbol{a^2 \bar{\mu}^2})} + {\alpha}_1 + {\alpha}_2 \ (\Red{\boldsymbol{{\omega}_{A_1}}} + \Magenta{\boldsymbol{{\omega}_{A_2}}}) + {\alpha}_3 \ (\Red{\boldsymbol{{\omega}_{A_1}^2}} + \Magenta{\boldsymbol{{\omega}_{A_2}^2}}) + {\alpha}_4 \ \Red{\boldsymbol{{\omega}_{A_1}}} \ \Magenta{\boldsymbol{{\omega}_{A_2}}} + {\alpha}_5 \ (\Red{\boldsymbol{{\omega}_{A_1}^3}} + \Magenta{\boldsymbol{{\omega}_{A_2}^3}}) \nonumber \\
&& + \ {\alpha}_6 \ \Red{\boldsymbol{{\omega}_{A_1}}} \ \Magenta{\boldsymbol{{\omega}_{A_2}}} \ (\Red{\boldsymbol{{\omega}_{A_1}}} + \Magenta{\boldsymbol{{\omega}_{A_2}}}) + {\alpha}_7 \ (\Red{\boldsymbol{{\omega}_{A_1}^4}} + \Magenta{\boldsymbol{{\omega}_{A_2}^4}}) + {\alpha}_8 \ \Red{\boldsymbol{{\omega}_{A_1}^2}} \ \Magenta{\boldsymbol{{\omega}_{A_2}^2}} + {\alpha}_9 \ \Red{\boldsymbol{{\omega}_{A_1}}} \ \Magenta{\boldsymbol{{\omega}_{A_2}}} \ (\Red{\boldsymbol{{\omega}_{A_1}^2}} + \Magenta{\boldsymbol{{\omega}_{A_2}^2}})\nonumber \\
&& + \ {\alpha}_{10} \ \Red{\boldsymbol{{\omega}_{A_1}^2}} \ \Magenta{\boldsymbol{{\omega}_{A_2}^2}} \ (\Red{\boldsymbol{{\omega}_{A_1}}} + \Magenta{\boldsymbol{{\omega}_{A_2}}}) + {\alpha}_{11} \ \Red{\boldsymbol{{\omega}_{A_1}}} \ \Magenta{\boldsymbol{{\omega}_{A_2}}} \ (\Red{\boldsymbol{{\omega}_{A_1}^3}} + \Magenta{\boldsymbol{{\omega}_{A_2}^3}}) + {\alpha}_{12} \ \Red{\boldsymbol{{\omega}_{A_1}^3}} \ \Magenta{\boldsymbol{{\omega}_{A_2}^3}} + {\alpha}_{13} \ \Red{\boldsymbol{{\omega}_{A_1}^2}} \ \Magenta{\boldsymbol{{\omega}_{A_2}^2}} \ (\Red{\boldsymbol{{\omega}_{A_1}^2}} + \Magenta{\boldsymbol{{\omega}_{A_2}^2}})\nonumber \\
&& + \ {\alpha}_{14} \ \Red{\boldsymbol{{\omega}_{A_1}^3}} \ \Magenta{\boldsymbol{{\omega}_{A_2}^3}} \ (\Red{\boldsymbol{{\omega}_{A_1}}} + \Magenta{\boldsymbol{{\omega}_{A_2}}}) + {\alpha}_{15} \ \Red{\boldsymbol{{\omega}_{A_1}^4}} \ \Magenta{\boldsymbol{{\omega}_{A_2}^4}} + {\alpha}_{16} \ (\Blue{\boldsymbol{{\omega}_{O_1}}} + \Cyan{\boldsymbol{{\omega}_{O_2}}}) + {\alpha}_{17} \ \Blue{\boldsymbol{{\omega}_{O_1}}} \ \Cyan{\boldsymbol{{\omega}_{O_2}}}\nonumber \\ 
&& + \ {\alpha}_{18} \ (\Red{\boldsymbol{{\omega}_{A_1}}} + \Magenta{\boldsymbol{{\omega}_{A_2}}}) \  (\Blue{\boldsymbol{{\omega}_{O_1}}} + \Cyan{\boldsymbol{{\omega}_{O_2}}}) + {\alpha}_{19} \ \Red{\boldsymbol{{\omega}_{A_1}}} \ \Magenta{\boldsymbol{{\omega}_{A_2}}} \ (\Blue{\boldsymbol{{\omega}_{O_1}}} + \Cyan{\boldsymbol{{\omega}_{O_2}}}) \nonumber \\
&& + \ {\alpha}_{20} \ \Big[(\Red{\boldsymbol{\omega_{A_1}^2}} + \Magenta{\boldsymbol{\omega_{A_2}^2}}) \ (\Blue{\boldsymbol{\omega_{O_1}}} + \Cyan{\boldsymbol{\omega_{O_2}}}) + (\Red{\boldsymbol{\omega_{A_1}}} + \Magenta{\boldsymbol{\omega_{A_2}}}) \ \Blue{\boldsymbol{\omega_{O_1}}} \ \Cyan{\boldsymbol{\omega_{O_2}}} \Big] + {\alpha}_{21} \ (\Red{\boldsymbol{\omega_{A_1}^2}} + \Magenta{\boldsymbol{\omega_{A_2}^2}}) \ \Blue{\boldsymbol{\omega_{O_1}}} \ \Cyan{\boldsymbol{\omega_{O_2}}}\nonumber \\
&& + \ {\alpha}_{22} \ (\Red{\boldsymbol{\omega_{A_1}^3}} + \Magenta{\boldsymbol{\omega_{A_2}^3}}) \ (\Blue{\boldsymbol{\omega_{O_1}}} + \Cyan{\boldsymbol{\omega_{O_2}}}) + {\alpha}_{23} \ \Red{\boldsymbol{\omega_{A_1}}} \ \Magenta{\boldsymbol{\omega_{A_2}}} \ \Big[ (\Red{\boldsymbol{\omega_{A_1}}} + \Magenta{\boldsymbol{\omega_{A_2}}}) \ (\Blue{\boldsymbol{\omega_{O_1}}} + \Cyan{\boldsymbol{\omega_{O_2}}}) + \Blue{\boldsymbol{\omega_{O_1}}} \ \Cyan{\boldsymbol{\omega_{O_2}}}\Big] \nonumber \\
&& + \ {\alpha}_{24} \ (\Red{\boldsymbol{\omega_{A_1}^3}} + \Magenta{\boldsymbol{\omega_{A_2}^3}}) \  \Blue{\boldsymbol{\omega_{O_1}}} \ \Cyan{\boldsymbol{\omega_{O_2}}} + {\alpha}_{25} \ \Red{\boldsymbol{\omega_{A_1}}} \ \Magenta{\boldsymbol{\omega_{A_2}}} \ (\Red{\boldsymbol{\omega_{A_1}^2}} + \Magenta{\boldsymbol{\omega_{A_2}^2}}) \ (\Blue{\boldsymbol{\omega_{O_1}}} + \Cyan{\boldsymbol{\omega_{O_2}}})\nonumber \\
&& + \ {\alpha}_{26} \ \Red{\boldsymbol{\omega_{A_1}}} \ \Magenta{\boldsymbol{\omega_{A_2}}} \ \Big[\Red{\boldsymbol{\omega_{A_1}}} \ \Magenta{\boldsymbol{\omega_{A_2}}} \ (\Blue{\boldsymbol{\omega_{O_1}}} + \Cyan{\boldsymbol{\omega_{O_2}}}) + (\Red{\boldsymbol{\omega_{A_1}}} + \Magenta{\boldsymbol{\omega_{A_2}}}) 
 \Blue{\boldsymbol{\omega_{O_1}}} \ \Cyan{\boldsymbol{\omega_{O_2}}} \Big] + {\alpha}_{27} \ \Red{\boldsymbol{\omega_{A_1}^2}} \ \Magenta{\boldsymbol{\omega_{A_2}^2}} \ \Blue{\boldsymbol{\omega_{O_1}}} \ \Cyan{\boldsymbol{\omega_{O_2}}} \nonumber \\
&& + \ {\alpha}_{28} \ \Red{\boldsymbol{\omega_{A_1}}} \ \Magenta{\boldsymbol{\omega_{A_2}}} \ \Big[\Red{\boldsymbol{\omega_{A_1}}} \ \Magenta{\boldsymbol{\omega_{A_2}}} \ (\Red{\boldsymbol{\omega_{A_1}}} + \Magenta{\boldsymbol{\omega_{A_2}}}) \ (\Blue{\boldsymbol{\omega_{O_1}}} + \Cyan{\boldsymbol{\omega_{O_2}}}) + (\Red{\boldsymbol{\omega_{A_1}^2}} + \Magenta{\boldsymbol{\omega_{A_2}^2}}) \ \Blue{\boldsymbol{\omega_{O_1}}} \ \Cyan{\boldsymbol{\omega_{O_2}}} \Big] \nonumber \\
&& + \ {\alpha}_{29} \ \Red{\boldsymbol{\omega_{A_1}^3}} \ \Magenta{\boldsymbol{\omega_{A_2}^3}} \ (\Blue{\boldsymbol{\omega_{O_1}}} + \Cyan{\boldsymbol{\omega_{O_2}}}) + {\alpha}_{30} \ \Red{\boldsymbol{\omega_{A_1}^2}} \ \Magenta{\boldsymbol{\omega_{A_2}^2}} \ (\Red{\boldsymbol{\omega_{A_1}}} + \Magenta{\boldsymbol{\omega_{A_2}}}) \ \Blue{\boldsymbol{\omega_{O_1}}} \ \Cyan{\boldsymbol{\omega_{O_2}}} + {\alpha}_{31} \ \Red{\boldsymbol{\omega_{A_1}^3}} \ \Magenta{\boldsymbol{\omega_{A_2}^3}} \ \Blue{\boldsymbol{\omega_{O_1}}} \ \Cyan{\boldsymbol{\omega_{O_2}}} \Bigg\rbrace + \Op (g_\circ^6) \nonumber 
\label{ZA_stag}
\eea
\end{small}
where $c_F \equiv (N_c^2 - 1)/(2 N_c)$ and $N_f$ is the number of flavors. The numerical constants $\alpha_i$ have been computed for various sets of values of the Symanzik coefficients; their values are listed in Table \ref{tab2} for the Wilson, tree-level (TL) Symanzik and Iwasaki gluon actions. In the $RI'$-alternative scheme, the above result is modified by adding the finite term $g_\circ^4 / (4 \pi)^4 \ c_F \ N_f$. Also, in the $\overline{MS}$ scheme we must add the finite term $g_\circ^4 / (4 \pi)^4 (- 3 / 2 \ c_F N_f)$.
\begin{table}[thb]
  \centering
  \begin{footnotesize}
  \begin{tabular}{llllllll}\toprule
  & Wilson & TL Symanzik & Iwasaki & & Wilson & TL Symanzik & Iwasaki  \\\midrule
${\alpha}_{1}$ & 17.420(1) & 16.000(1) & 14.610(1)     & ${\alpha}_{16}$ & 24.9873(2) & 18.0489(4) & 9.9571(2) \\
${\alpha}_{2}$ & -116.049(7) & -81.342(5) & -41.583(2)   & ${\alpha}_{17}$ & -97.4550(2) & -62.2675(1) & -26.5359(1) \\
${\alpha}_{3}$ & 839.788(9) & 539.121(6) & 230.050(1)    & ${\alpha}_{18}$ & -292.3650(5) & -186.8025(4) & -79.6078(2) \\
${\alpha}_{4}$ & 2175.14(3) & 1394.12(2) & 591.88(1) & ${\alpha}_{19}$ & 4864.513(9) & 2921.876(6) & 1107.333(2) \\
${\alpha}_{5}$ & -3462.830(1) & -2098.136(5) & -801.633(3) & ${\alpha}_{20}$ & 1621.504(3) & 973.959(2) & 369.111(1) \\
${\alpha}_{6}$ & -19565.9(1) & -11858.6(1) & -4528.6(1) & ${\alpha}_{21}$ & -10617.81(2) & -6122.11(1) & -2169.30(1) \\
${\alpha}_{7}$ & 6424.33(2) & 3740.18(1) & 1337.93(1)    & ${\alpha}_{22}$ & -3539.269(6) & -2040.705(4) & -723.099(1) \\
${\alpha}_{8}$ & 200966.5(4) & 117179.7(4) & 41977.1(1)   & ${\alpha}_{23}$ & -31853.42(5) & -18366.34(3) & -6507.89(1) \\
${\alpha}_{9}$ & 92171.5(3) & 53720.8(1) & 19237.6(1)      & ${\alpha}_{24}$ & 25847.14(3) & 14435.59(2) & 4881.52(1) \\
${\alpha}_{10}$ & -1026448(1) & -580271(2) & -198722(1)      & ${\alpha}_{25}$ & 77541.41(1) & 43306.78(6)  & 14644.54(2) \\
${\alpha}_{11}$ & -183998.3(3) & -103929.7(3) & -35561.1(1) & ${\alpha}_{26}$ & 232624.2(3) & 129920.3(2) & 43933.6(1) \\
${\alpha}_{12}$ & 5517230(30) &  3037110(10) & 1003641(1)     & ${\alpha}_{27}$ & -1844375(1) & -1002465(1) & -326727(1) \\
${\alpha}_{13}$ & 2145810(10) & 1180684(4) & 389979(1)   & ${\alpha}_{28}$ & -614791.6(6) & -334155.0(4) & -108909.0(2) \\
${\alpha}_{14}$ & -11889300(40) & -6386950(30) & -2046240(10) & ${\alpha}_{29}$ & 1736048.1(8) & 920956.7(7) & 290916.1(3) \\
${\alpha}_{15}$ & 26137700(200) & 13729010(10) & 4278680(10) & ${\alpha}_{30}$ & 5208144(2) & 2762870(2) & 872748(1) \\
                &            &            &            & ${\alpha}_{31}$ & -15545543(1) & -8065557(2) & -2478207(1) \\\bottomrule
  \end{tabular}
  \end{footnotesize}
  \caption{Numerical coefficients for the Axial Vector operator.}
\label{tab2}
\end{table}

\begin{figure}[thb] 
  \centering
  \includegraphics[width=4cm,clip]{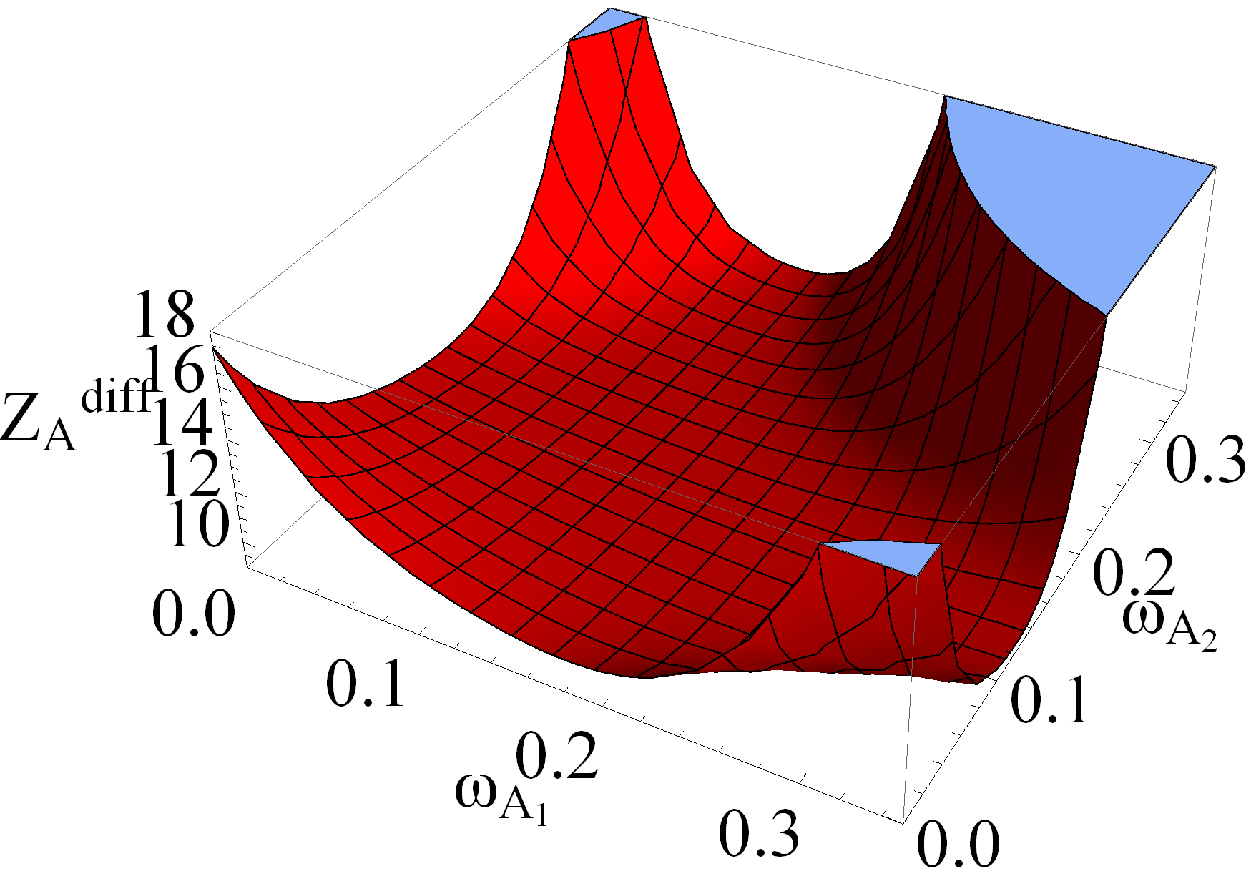} \hspace{0.1cm}
  \includegraphics[width=4cm,clip]{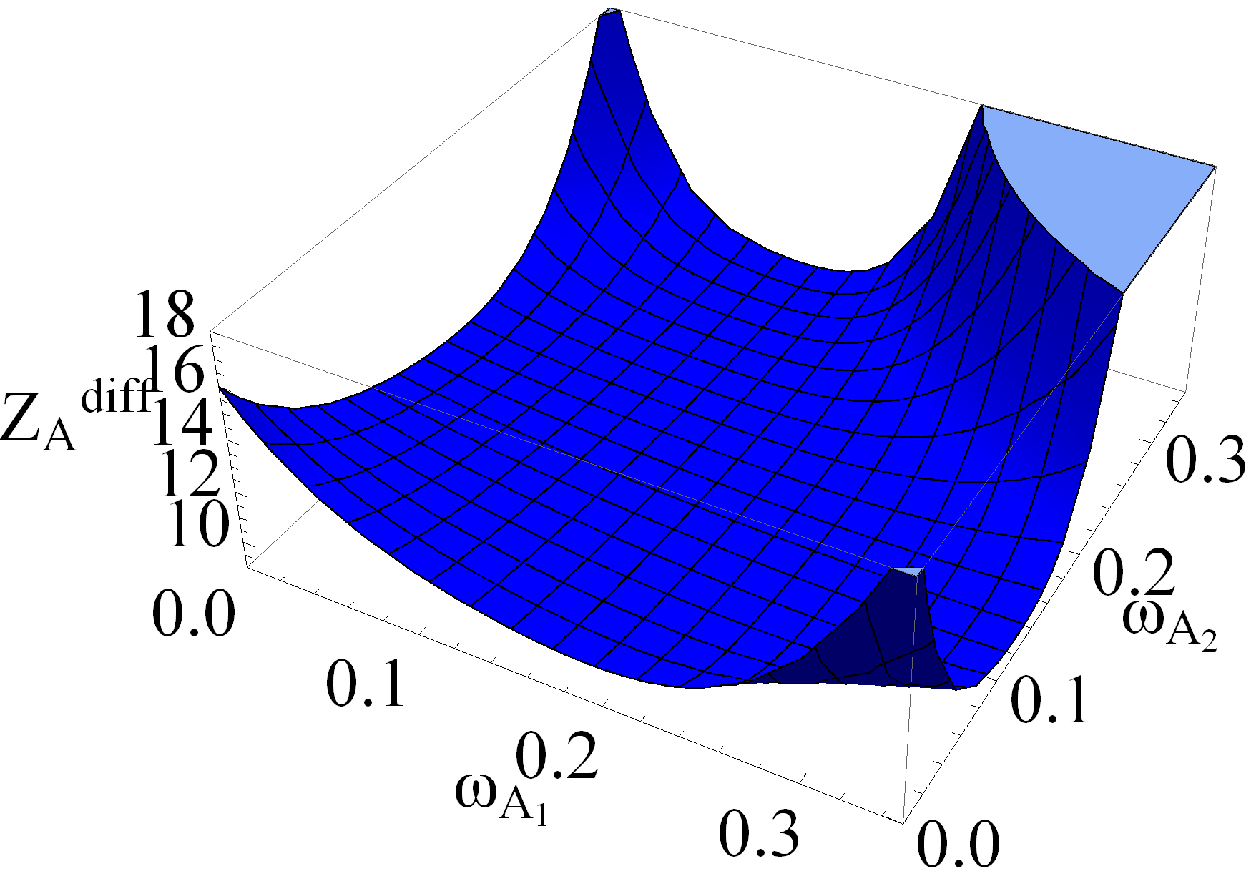} \hspace{0.1cm}
  \includegraphics[width=4cm,clip]{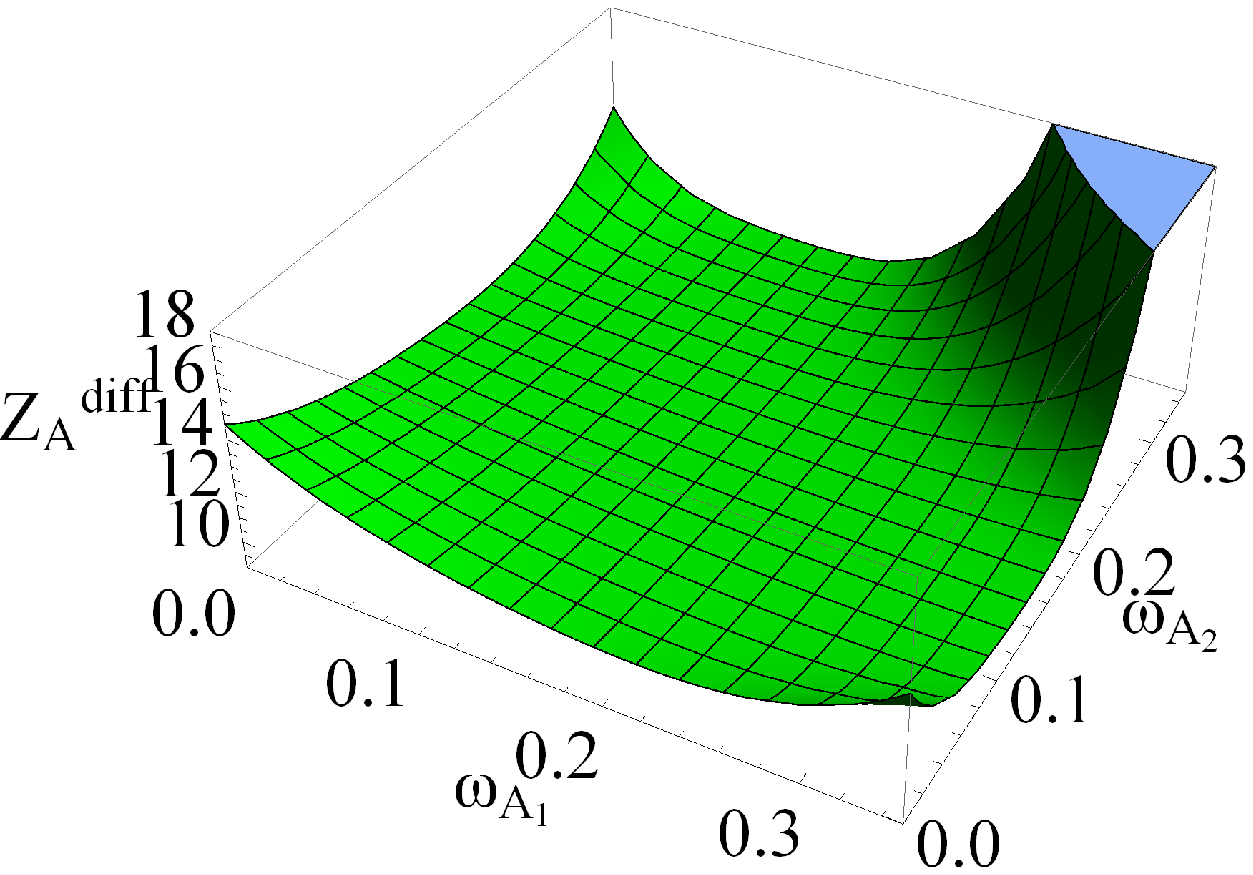}
  \caption{Plots of $Z_A^{\rm{diff.}}$, as a function of $\omega_{A_1}$ and $\omega_{A_2}$ for $\omega_{O_1} = \omega_{O_2} = 0$ $ \ $ (left: Wilson action, center: TL Symanzik action, right: Iwasaki action).}
\label{Stagplots3D1}
\end{figure}

\begin{figure}[thb] 
  \centering
  \includegraphics[width=4cm,clip]{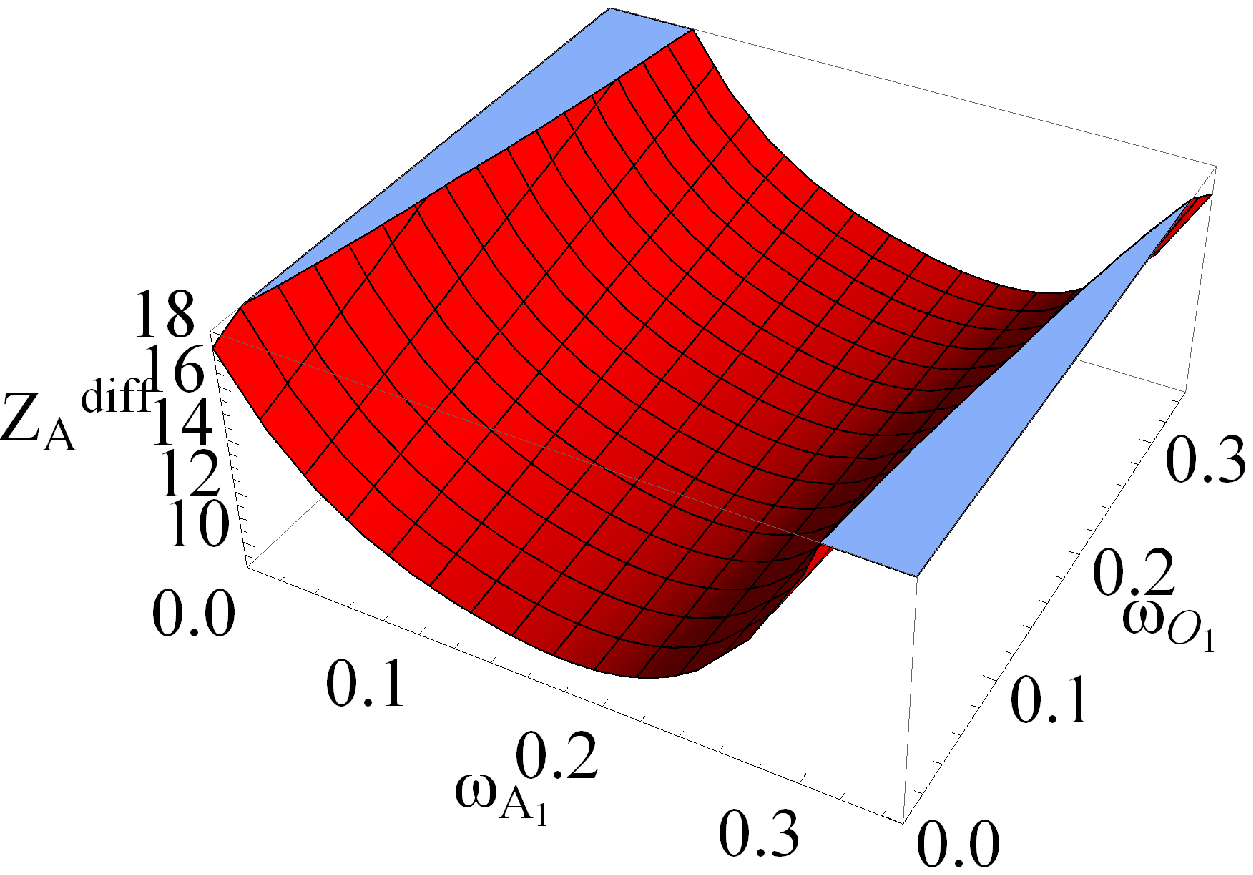} \hspace{0.1cm}
  \includegraphics[width=4cm,clip]{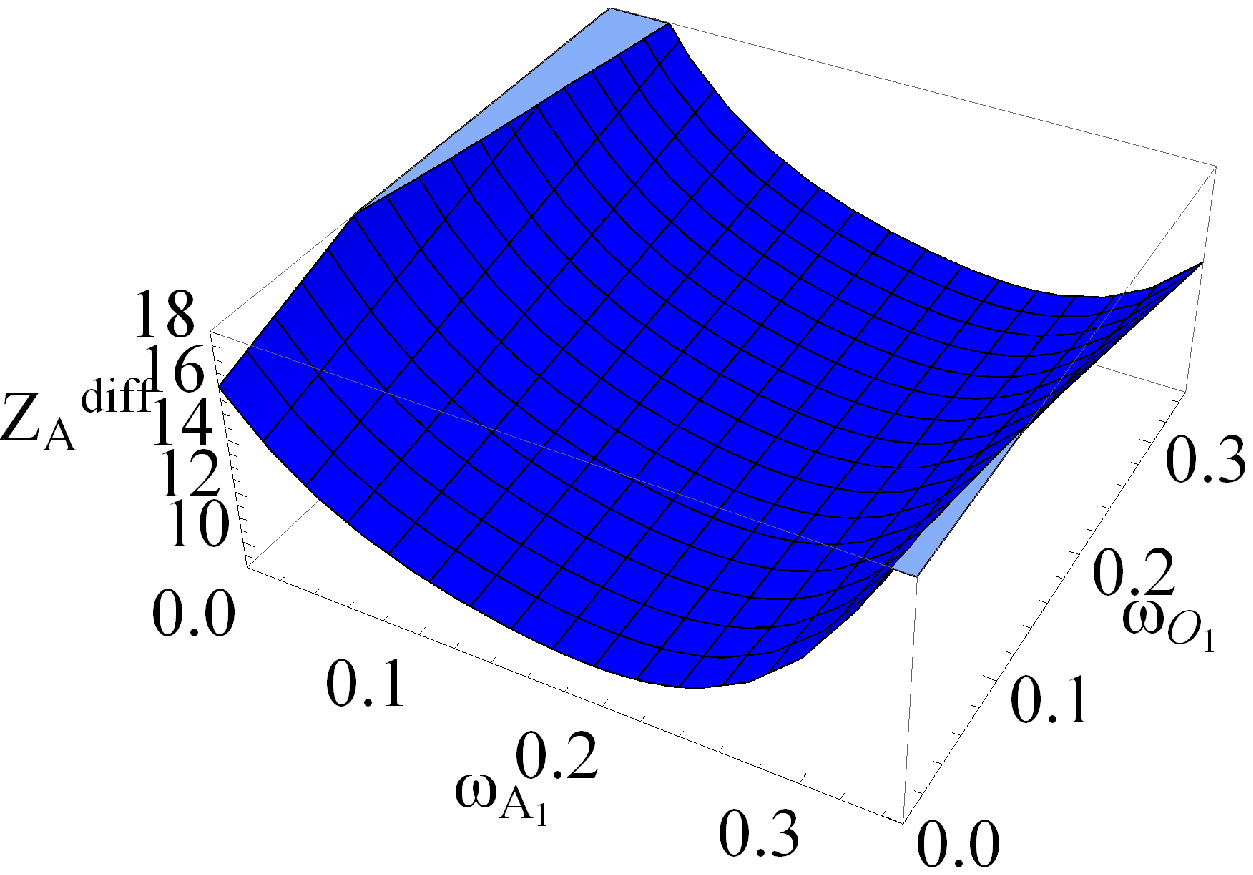} \hspace{0.1cm}
  \includegraphics[width=4cm,clip]{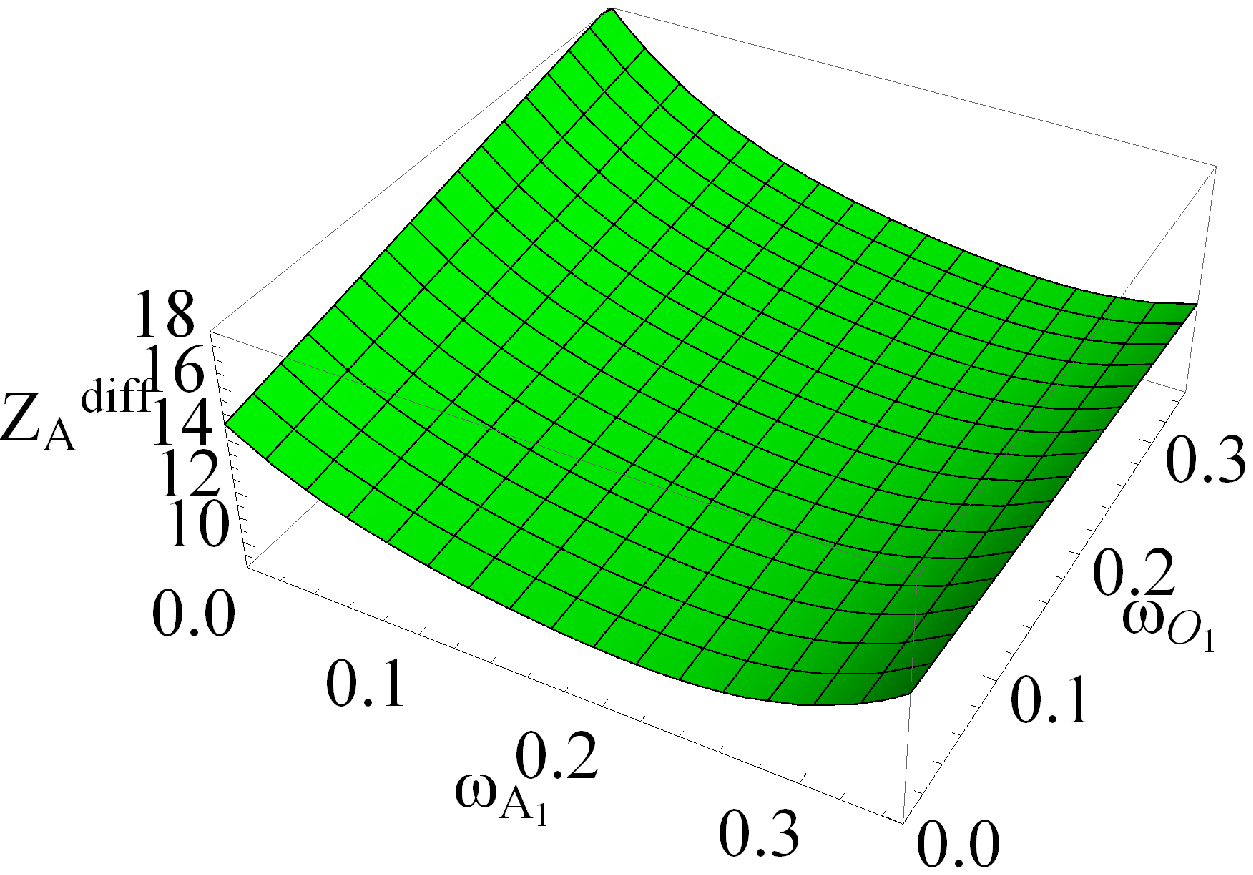}
  \caption{Plots of $Z_A^{\rm{diff.}}$, as a function of $\omega_{A_1}$ and $\omega_{O_1}$ for $\omega_{A_2} = \omega_{O_2} = 0$ $ \ $ (left: Wilson action, center: TL Symanzik action, right: Iwasaki action).}
\label{Stagplots3D2}
\end{figure}

In Figs. \ref{Stagplots3D1}-\ref{Stagplots3D2} we illustrate our result in the $RI'$ scheme by selecting certain values of $c_i, \ \omega_{A_1}, \ \omega_{A_2}, \ \omega_{O_1}$ and $\omega_{O_2}$. The vertical axis of these plots corresponds to $Z_A^{\rm{diff.}} \equiv \left[ Z_A^{\rm{(singlet)}}\! \left( a \bar{\mu} \right) - Z_A^{\rm{(nonsinglet)}}\! \left( a \bar{\mu} \right) \right] \left( - \frac{g_o^4}{\left( 4 \pi \right)^4} N_f c_F \right)^{-1}$ for $\bar{\mu} = 1 / a$. We notice that the plots for the Iwasaki action are flatter than the remaining actions. Also, in Fig. \ref{Stagplots3D1} we notice that there is only one minimum, on the $45^{\circ}$ axis. Therefore, the two smearing steps of the fermion action give better results than only one smearing step. Furthermore, we observe that the stout smearing of the action is more effective in minimizing $Z_A^{\rm{diff.}}$ than the stout smearing of operators. Some other graphs of our result for certain values of the Symanzik coefficients and the stout smearing parameters can be found in \cite{Constantinou:2016ieh}.
\vspace{-0.69cm}
\vspace{-0.7cm}
\bibliography{lattice2017}

\begin{thebibliography}{18}

\bibitem{Constantinou:2016ieh}
M.~Constantinou et~al., Phys. Rev. \textbf{D94}, 114513 (2016),
  \texttt{[arXiv:1610.06744]}

\bibitem{QCDSF:2011aa}
G.S. Bali et~al., Phys. Rev. Lett. \textbf{108}, 222001 (2012),
  \texttt{[arXiv:1112.3354]}

\bibitem{Chambers:2014pea}
A.J. Chambers et~al. (QCDSF), Phys. Lett. \textbf{B740}, 30 (2015),
  \texttt{[arXiv:1410.3078]}

\bibitem{Bouchard:2016heu}
C.~Bouchard et~al., Phys. Rev. \textbf{D96}, 014504 (2017),
  \texttt{[arXiv:1612.06963]}

\bibitem{Bali:2017jyw}
G.S. Bali et~al., PoS \textbf{LATTICE2016}, 187 (2016),
  \texttt{[arXiv:1703.03745]}

\bibitem{Constantinou:2014rka}
M.~Constantinou et~al., PoS \textbf{LATTICE2014}, 298 (2014),
  \texttt{[arXiv:1411.6990]}

\bibitem{Aoki:2005vt}
Y.~Aoki, Z.~Fodor, S.D. Katz, K.~Szabo, JHEP \textbf{0601}, 089 (2006),
  \texttt{[hep-lat/0510084]}

\bibitem{Borsanyi:2011bm}
Bors\'anyi et~al., J. Phys. \textbf{G38}, 124060 (2011),
  \texttt{[arXiv:1109.5030]}

\bibitem{Bazavov:2012zad}
A.~Bazavov et~al. (2012), \texttt{[arXiv:1212.4768]}

\bibitem{Constantinou:2013pba}
M.~Constantinou et~al., Phys. Rev. \textbf{D88}, 034504 (2013),
  \texttt{[arXiv:1305.1870]}

\bibitem{Morningstar:2003gk}
C.~Morningstar, M.J. Peardon, Phys. Rev. \textbf{D69}, 054501 (2004),
  \texttt{[hep-lat/0311018]}

\bibitem{Horsley:2004mx}
R.~Horsley et~al., Nucl. Phys. \textbf{B693}, 3 (2004), [Erratum: Nucl. Phys.
  B713, 601(2005)], \texttt{[hep-lat/0404007]}

\bibitem{Patel:1992vu}
A.~Patel, S.R. Sharpe, Nucl. Phys. \textbf{B395}, 701 (1993),
  \texttt{[hep-lat/9210039]}

\bibitem{Gracey:2003yr}
J.A. Gracey, Nucl. Phys. \textbf{B662}, 247 (2003), \texttt{[hep-ph/0304113]}

\bibitem{Larin:1993tq}
S.A. Larin, Phys. Lett. \textbf{B303}, 113 (1993), \texttt{[hep-ph/9302240]}

\bibitem{Luscher:1995np}
M.~L{\"u}scher, P.~Weisz, Nucl. Phys. \textbf{B452}, 234 (1995),
  \texttt{[hep-lat/9505011]}

\bibitem{Chetyrkin:1981}
K.~Chetyrkin, F.~Tkachov, Nucl. Phys. \textbf{B192}, 159 (1981)

\bibitem{Panagopoulos:1990}
H.~Panagopoulos, E.~Vicari, Nucl. Phys. \textbf{B332}, 261 (1990)

\end{thebibliography}

\end{document}